\documentstyle[12pt,fleqn]{article}
\begin{document}                                         
\sloppy
 
\def\beq{\begin{equation}}
\def\eeq{\end{equation}}
\newcommand{\av}[1]{\langle #1\rangle}

\vspace*{20mm}
\begin{center}

{\Large Infinite matrices may violate the associative law}\vfill

Ofir E. Alon \ and \ Nimrod Moiseyev\medskip

{\em Department of Chemistry, Technion---Israel Institute of Technology,
32 000 Haifa, Israel}\medskip

and\medskip

Asher Peres\medskip

{\em Department of Physics, Technion---Israel Institute of Technology,
32 000 Haifa, Israel}\vfill

{\bf Abstract}\end{center}

\begin{quote} The momentum operator for a particle in a box is
represented by an infinite order Hermitian matrix $P$. Its square $P^2$
is well defined (and diagonal), but its cube $P^3$ is ill defined,
because $P\,P^2\neq P^2\,P$. Truncating these matrices to a finite order
restores the associative law, but leads to other curious
results.\end{quote}\vfill

\noindent Classification \ 0260 \ (02.10.Sp) \vfill\newpage

Matrices of infinite order are used in quantum mechanics for
representing dynamical variables. It is commonly assumed that these
matrices obey the usual laws valid for finite matrices. However, it is
obvious that difficulties may occasionally arise because of divergent
infinite sums. We give here a very simple example of a Hermitian matrix
$P$ such that $P^2$ is well defined, but $P^3$ is not, because
$P\,P^2\neq P^2\,P$.

Related ambiguities may also arise in numerical calculations, when
infinite matrices are truncated to finite size (because there are no
infinite computers). While the associative law is obviously fulfilled,
the difficulty emerges in some other way. In particular, we sometimes
encounter in quantum mechanics operators which are symmetric, but are
not self-adjoint (for example, the radial momentum operator). The
representation of these operators by Hermitian matrices conceals their
lack of self-adjointness and leads to curious results, as in the
following example.

Consider the Hilbert space of functions $f(x)$, with $0\leq x\leq\pi$,
and inner product

\beq \av{f,g}=\int^\pi_0 f^*(x)\,g(x)\,dx. \eeq
In that space, the set of functions $u_m(x)=(2/\pi)^{1/2}\sin mx$, for
all positive integers $m$, can be taken as a complete orthonormal basis
(namely, any square integrable function of $x$ can be written as a
linear combination of the $u_m$, except at a finite number of isolated
points). The operator $p=-id/dx$ has matrix elements

\beq P_{mn}=\int^\pi_0 u_m\,(-id/dx)\,u_n\,dx, \eeq
\[ \phantom{P_{mn}}=\left\{\begin{array}{lll}
  -4imn/\pi(m^2-n^2) & \quad & {\rm if}\quad m+n\quad {\rm is\ odd},\\
  0 & & {\rm if}\quad m+n\quad {\rm is\ even}.\end{array}\right.\]
This operator, currently defined only on functions which vanish at
$x=0$ and $x=\pi$, has the physical meaning of linear momentum of a
particle in a rigid one-dimensional box.  It can be directly verified
by algebraic methods~[1] that

\beq \sum_{s=1}^\infty P_{ms}\,P_{sn}=mn\,\delta_{mn}\,, \eeq
as could be expected from

\beq (P^2)_{mn}=\int^\pi_0 u_m\,(-id/dx)^2\,u_n\,dx. \eeq

However, difficulties occur if we try to define likewise

\beq (P^3)_{mn}\stackrel{?}{=}\int^\pi_0 u_m\,(-id/dx)^3\,u_n\,dx. \eeq
The result is not Hermitian. This should not be a surprise, because,
with the above boundary conditions, the operator $-id/dx$ is
not self-adjoint~[2,3]. The matrices $P_{mn}$ and $(P^2)_{mn}$ in
Eqs.~(2) and (4) were Hermitian, because it was then possible to perform
integrations by parts, in which the boundary terms vanished. However, if
we attempt to do the same with Eq.~(\theequation), we obtain

\beq \int_0^\pi u'_m\,u''_n\,dx=-\int_0^\pi u''_m\,u'_n\,dx+
  \Bigl[u'_m\,u'_n\Bigr]^\pi_0\,, \eeq
and the last term does not vanish.

The source of the difficulty is that the domain of $P$ consists of
functions on the interval $[0,\pi]$ which vanish at the interval
endpoints. However, $Pu_n$ does not vanish at these endpoints, and
therefore does not belong to the domain of $P$: the expression $P(Pu_n)$
is not mathematically defined. This does not contradict the fact that
$(-d^2/dx^2)u_n=n^2u_n$. The point is that, when $P$ is defined as
above, $P^2$ is not the same as $(-d^2/dx^2)$, notwithstanding Eq.~(4).
These two operators coincide in the common part of their domains of
definition, but the domain of definition of $P^2$ is smaller.

The issue we are investigating here is how these curious properties
appear when matrix notations are used for representing differential
operators (for example in a mundane numerical analysis).  If we try to
use algebraic methods for defining $P^3$, we encounter the same
difficulty in another form:

\beq \sum_{s=1}^\infty P_{ms}\,(P^2)_{sn}\neq
     \sum_{s=1}^\infty (P^2)_{ms}\,P_{sn}\,. \eeq
Let us examine why the associative law fails. In the present case, the
right hand side of

\beq
(P^3)_{mn}\stackrel{?}{=}\sum_{r,s}^\infty\,P_{mr}\,P_{rs}\,P_{sn}\,,
\eeq
contains infinitely many terms with $r=s\pm 1\gg m+n$, which
behave as $\pm\sum 1/r$. Therefore the sum of positive terms diverges,
the sum of negative terms diverges, and the entire sum in (\theequation)
is only conditionally convergent. Its value depends on the order of
summation.

It is interesting to see how this difficulty is reflected in numerical
calculations, where infinite matrices are truncated and replaced by
finite matrices of order $N$ (some large number). Instead of
(\theequation), we may try to define

\beq
S_{mn}=\lim_{N\to\infty}\,\sum_{r,s}^N\,P_{mr}\,P_{rs}\,P_{sn}\,, \eeq
and the question is whether this sum indeed tends to a unique limit as
$N\to\infty$. In the present case, it does, as shown below. However,
this happens only because of delicate cancellations between positive and
negative terms, which would not occur in general.

Consider for definiteness the case where $m$ and $s$ are odd, and $n,\
r$, and $N$ are even. The question is whether, for $N\gg m+n$, the
contributions of $r=N$ and $s=N-1$ to the sum in Eq.~(\theequation) are
vanishingly small. These contributions are, apart from an overall factor
$64imn/\pi^3$,

\beq
\frac{N^2}{m^2-N^2}\,\sum_{s=1}^{N-1}\frac{s^2}{(N^2-s^2)\,(s^2-n^2)}+
\frac{(N-1)^2}{(N-1)^2-n^2}\,\sum_{r=2}^{N-2}
\frac{r^2}{(m^2-r^2)\,[r^2-(N-1)^2]}\,. \eeq
(The second sum runs only to $r=N-2$, in order to avoid double counting
of the $r,s=N,N\!-\!1$ matrix element.) In the above sums, the main
contribution comes from terms where $r$ and $s$ are close to $N$, so
that the various terms are of the order of $N^{-1}$, rather than
$N^{-2}$. We therefore write $r=N-2k$ and $s=N-1-2k$, and we neglect
terms with $m^2$ and $n^2$, which are much smaller than $N^2$. The two
sums in (\theequation) become

\beq -\,\sum_{k=0}\frac{1}{(2k+1)\,(2N-2k-1)}+
  \sum_{k=1}\frac{1}{(2k-1)\,(2N-2k-1)}\,. \eeq
To be consistent with the preceding approximation, the sums in
(\theequation) run up to values of $k$ which are much smaller than $N/2$
(so that both $r$ and $s$ are close to $N$). We thus obtain,
approximately,

\beq  \frac{1}{2N}\biggl(-\,\sum_{k=0}\frac{1}{2k+1}+
   \sum_{k=1}\frac{1}{2k-1}\biggr)={1\over N}\biggl({1\over 2}+
   \sum_{k=0}\frac{1}{4k^2-1}\biggr). \eeq
For large $N$, the sums on the left hand side behave as $\mp\log N$.
However, the sum on the right hand side tends to $-{1\over 2}$ (see
ref.~[1]\,) so that the whole expression vanishes. In summary, the total
contribution of the $N-1$ and $N$ terms to the sum in Eq.~(9) is of the
order of $N^{-2}$, and that sum converges (for the particular matrix
$P_{mn}$ under consideration).

We have found empirically that, as $N$ becomes much larger than $m$ and
$n$ (which are kept fixed), matrix elements $S_{mn}$ tends to the
Hermitian part of the right hand side of Eq.~(5), namely to

\beq  R_{mn}=(i/2)\,\biggl[\int_0^\pi u_m\,(d/dx)^3\,u_n\,dx
   -\int_0^\pi u_n\,(d/dx)^3\,u_m\,dx\biggr]. \eeq
Note that the right hand side of (\theequation) can also be written,
after integration by parts, as\clearpage

\[ (i/2)\,\int_0^\pi(-u'_m\,u''_n+u'_n\,u''_m)\,dx=
 (i/2)\,\int_0^\pi(n^2\, u'_m\,u_n-m^2\,u'_n\,u_m)\,dx, \]\vspace{-4mm}
\beq \hspace*{30mm}=(m^2\,P_{mn}-n^2\,P_{nm})/2=
  (m^2\,P_{mn}+P_{mn}\,n^2)/2, \eeq
which is the average of $P^2P$ and $PP^2$. The convergence was found to
be of oscillatory type, as illustrated in Table 1, for a few randomly
chosen matrix elements.

\begin{center}
Table 1. \ Convergence of $S_{mn}$ toward $R_{mn}$ (the table lists
values of $-iS_{mn}$).\\[2mm]
\begin{tabular}{ccccccccc}\hline\hline
$m$ & $n$ & $N\!=\!99$ & $N\!=\!100$ & $N\!=\!999$ & $N\!=\!1000$ &
$N\!=\!1999$ & $N\!=\!2000$ & $-iR_{mn}$ \\ \hline 
1 & 2 & 2.156 & 2.088 & 2.127 & 2.117 & 2.125 & 2.120 & 2.122 \\ 
2 & 3 & 9.828 & 10.032 & 9.918 & 9.945 & 9.924 & 9.939 & 9.931 \\ 
20 & 31 & 935.67 & 959.59 & 956.05 & 958.89 & 956.77 & 958.31 & 957.56 
\\ 
60 & 91 & 5198.7 & 6667.9 & 8803.1 & 8828.4 & 8814.0 & 8827.5 & 8822.4
\\ \hline\hline \end{tabular} \end{center}

On the other hand, truncation may also lead to counterintuitive results.
For example, if we define

\beq Q_{mn}=\left\{\begin{array}{llcl}
  P_{mn} &\quad{\rm if} & m\ {\rm and}\ n & \leq N, \\
  0 &\quad{\rm if} & m\ {\rm or}\ n & > N, \end{array}\right.\eeq
the eigenvalues of $Q_{mn}$ appear in opposite pairs, $\lambda$ and
$-\lambda$ (because $Q$ and $Q^T=-Q$ have the same eigenvalues). The
only exception occurs if $N$ is odd: the eigenvalue~0 is nondegenerate.
Therefore the eigenvalues of $Q^2$ are doubly degenerate (except the
null eigenvalue, if $N$ is odd). This should be contrasted with the
spectrum of $P^2$ which is 1, 4, 9, 16, \ldots

We have found empirically that the eigenvalues of $Q_{mn}$ are very
close to integers with a parity {\em opposite\/} to that of $N$. The
lowest and highest eigenvalues of $Q^2$ are listed in Table~2, for
$N=999$ and 1000. They appear quite different from those of $P^2$.
However, if we first compute $Q^2$ with $N=1000$, and then truncate it
by removing the 1000th row and column (or more rows and columns with the
largest indices), the eigenvalues of the resulting matrix are
nondegenerate and are very close to those of $P^2$.\clearpage

\begin{center}
Table 2. \ Some eigenvalues of $Q^2$, for $N=999$ and 1000.\\[2mm]
\begin{tabular}{rccc}\hline\hline
 & $N=999$ & $N=1000$ & $N=1000$ \\
 & (complete matrix) & (truncated to 999) & (complete matrix) \\ \hline 
1 & 0.000000 & 0.996663 & 0.996663 \\ 
2 & 3.986641 & 3.986641 & 0.996663 \\ 
3 & 3.986641 & 8.969969 & 8.969969 \\ 
4 & 15.94656 & 15.94656 & 8.969969 \\ 
5 & 15.94656 & 24.91658 & 24.91658 \\ 
$\cdots$ & $\cdots$ & $\cdots$ & $\cdots$ \\
996 & 988294.4 & 988294.4 & 986110.2 \\
997 & 988294.4 & 990283.2 & 990283.2 \\
998 & 992673.3 & 992673.3 & 990283.2 \\
999 & 992673.3 & 994666.5 & 994666.5 \\
1000 & & & 994666.5 \\ \hline\hline \end{tabular} \end{center}

These curious results can be explained by rearranging the rows and
columns of these matrices according to the sequence 1, 3,
\ldots\,,$N-1$, 2, 4, \ldots\,,$N-2$ or $N$. Since the only nonvanishing
matrix elements $(Q^2)_{mn}$ are those whose $m$ and $n$ have the same
parity, the matrices $Q^2$ separate into two blocks. In each matrix,
these two blocks have the same eigenvalues (so that each eigenvalue
appears twice, except the null eigenvalue, if $N$ is odd). It can be
seen by direct inspection that the truncated matrix also consists of two
blocks. Its odd-odd block is identical to the odd-odd block of
$Q^2_{(1000)}$\,, and its even-even block is the same as the even-even
block of $Q^2_{(999)}$\,. This explain why the same eigenvalue appears
in three different positions in Table~2. We have also found empirically
that removing additional rows and columns from $Q^2$ (those with the
largest indices) improves the convergence toward the ``true''
eigenvalues, given by Eq.~(3).

As higher powers of $P$ are considered, the results become curiouser and
curiouser. For instance, $P^4=P^2P^2$ is well defined, and has matrix
elements $m^2n^2\delta_{mn}$\,, by virtue of Eq.~(3). On the other hand,
$PP^2P$ is ill-defined. If we write

\beq (P\,P^2\,P)_{mn}=-\,\frac{16}{\pi^2}\,mn\,\sum_s\frac{s^4}
  {(m^2-s^2)\,(s^2-n^2)}\,, \eeq
where $s$ runs over all integers with a parity opposite to that of $m$
and $n$, {\em each\/} term in this sum tends to $-1$ when $s\to\infty$,
so that the sum is not even conditionally convergent. Moreover, if we
introduce truncated matrices, as in Eq.~(15), the sum
$\sum Q_{mr}\,Q_{rs}\,Q_{st}\,Q_{tn}$ does {\em not\/} tend to
$m^2n^2\delta_{mn}$\,. It diverges, for any finite $m$ and $n$ (with
same parity), when $N\to\infty$. The point is that $\sum Q_{mr}\,Q_{rs}$
converges to $ms\delta_{ms}$ only if $s$ is kept fixed as $N\to\infty$,
and likewise for $\sum Q_{st}\,Q_{tn}$\,. On the other hand, when we sum
over $s$ to obtain $(Q^4)_{mn}$\,, that sum includes terms where $s$ is
of the order of $N$ and the latter have a divergent contribution.

The conclusion to be drawn from these results is that truncation
methods, which are common practice in quantum mechanical calculations,
should be used with extreme caution when the truncated matrices
represent unbounded operators. In the particular case discussed here,
where the matrix $P$ is defined by Eq.~(2), the truncated matrix $P^2$
should be defined by squaring the truncated $P$ and then deleting the
last row and column (otherwise, the eigenvalues turn out to be
completely wrong). Likewise, the truncated matrix $P^3$ should be
defined by Eq.~(9), and {\em not\/} by Eq.~(5) which gives a
non-Hermitian result. However, these recipes may not be valid in
general, and other unbounded operators ought to be considered on a case
by case basis.\bigskip

Work by OEA was supported by the Technion Graduate School. Work by NM
was supported by the Basic Research Foundation administrated by the
Israel Academy of Sciences and Humanities, and by the Fund for
Encouragement of Research. Work by AP was supported by the Gerard Swope
Fund and by the Fund for Encouragement of Research.\bigskip

\begin{enumerate}
\item Prudnikov A P, Brychkov Yu A, and Marichev O I 1986 {\em Integrals
and Series\/} (New York: Gordon and Breach) Vol.~1, pp.~685--9
\item Riesz F and Sz.-Nagy B 1955 {\em Functional Analysis\/} (New York:
Ungar) p.~309
\item Peres A 1993 {\em Quantum Theory: Concepts and Methods\/}
(Dordrecht: Kluwer) p.~87
\end{enumerate}
\end{document}